\documentclass[fleqn,twoside]{article}
\usepackage{espcrc2}

\usepackage{epsfig}
\usepackage[figuresright]{rotating}

\newcommand{\ep}{\varepsilon}

\newcommand{\Li}[2]{{\mbox{Li}}_{#1}\left(#2\right)}
\newcommand{\Cl}[2]{{\mbox{Cl}}_{#1}\left(#2\right)}
\newcommand{\Ls}[2]{{\mbox{Ls}}_{#1}\left(#2\right)}
\newcommand{\LS}[3]{{\mbox{Ls}}_{#1}^{(#2)}\left(#3\right)}
\newcommand{\Lsc}[2]{{\mbox{Lsc}}_{#1\!}\left(#2\right)}
\newcommand{\tfrac}[2]{{\textstyle{\frac{#1}{#2}}}}

\newcommand{\Snp}[2]{{\it{S}}_{#1\!}\left(#2\right)}
\newcommand{\LsLsc}[4]{{\mbox{LsLsc}}_{#1,#2,#3} \left(#4\right)}

\newcommand{\AmS}{{\protect\the\textfont2
  A\kern-.1667em\lower.5ex\hbox{M}\kern-.125emS}}

\title{``Sixth root of unity'' and Feynman diagrams: hypergeometric function approach point of view}

\author{M.Yu. Kalmykov\address[MCSD]{
{\normalsize II. Institut f\"ur Theoretische Physik, Universit\"at Hamburg,} \\
{\normalsize Luruper Chaussee 149, 22761 Hamburg, Germany} }
        B.A. Kniehl\addressmark[MCSD]
}

\begin{document}

\begin{abstract}
We briefly discuss the transcendental constants generated 
through the $\ep$ expansion of generalized hypergeometric functions and their interrelation 
with the ``sixth root of unity.''
\vspace{1pc}
\end{abstract}

\maketitle

{\bf 1.}
After the appearance of the proof of the theorem concerning the calculability of 
4-loop renormalization group (RG) functions in the framework of dimensional regularization \cite{dimreg} 
in arbitrary renormalizable models in terms of $\zeta$ functions \cite{ibp} 
and the explicit evaluation of some 4- and 5-loop RG functions \cite{RG},
quite intriguing results were derived by Broadhurst \cite{MZV},
who had observed the appearance of non-zeta terms in the higher-order terms of the
$\ep$ expansion of the two-loop massless propagator diagram. 
Later, this observation was clearly explained in the framework of the knot approach to 
Feynman diagrams \cite{BK}, 
and all new transcendentals were parametrized in terms of multiple zeta values: 
$
\zeta_{\vec{s}} = 
\sum_{n_1>n_2> \ldots >n_k>0}
\prod_{j=1}^k \frac{1}{n_j^{s_j}}.
$
This result was extended to the case of the 2-loop massless propagator diagram in \cite{massless:2} 
and was recently fully analyzed in \cite{massless:full}.
In particular, it was shown that the $\ep$ expansions of 4-loop non-planar massless diagrams 
may generate transcendentals expressible in terms of multiple polylogarithms of 
the ``sixth root of unity.'' 
All recent results for the first few coefficients of the $\ep$ expansions of massless propagator diagrams \cite{last}
are in full agreement with this theorem. 
Let us recall that the appearance in Feynman diagrams of transcendental constants related to the ``sixth root of unity'' by
$
\sum_{n_1>n_2> \ldots >n_k>0}
\prod_{j=1}^k \frac{ \left( e^{{\rm i} \pi/3} \right)^{p_j n_j}}{n_j^{s_j}} \;,
$
where $p_j\in\{0,1,\cdots,5\}$,  
were predicted by Broadhurst in \cite{B99} in the context of the study of the finite parts of massive three-loop bubble diagrams. 
The lowest-weight Broadhurst bases include the following elements: 
$\pi^i \ln^j 2 \ln^k 3$, 
$\Cl{2}{\tfrac{\pi}{3}} \pi^i \ln^j 2 \ln^k 3$,  
$\Li{2}{\tfrac{1}{4}} \pi^i \ln^j 2 \ln^k 3$,  
etc., where $i,j,k$ are integers
(in \cite{B99} only constants up to  weight ${\bf 3}$ were presented). 
Based on the existing  partial results \cite{example},
the authors of \cite{FK99} pointed out that, in real physical diagrams 
with two massive cuts only, the modified set of transcendental constants 
differ as follows: (i) the factor $\frac{1}{\sqrt{3}}$
should be present (for example, $\pi, \ln 2, \ln 3$
of \cite{B99} should be  $\frac{\pi}{\sqrt{3}}, \ln 3$ as in \cite{FK99}); 
(ii)
all elements are generated by values of $\Li{a}{\exp\{i \theta_k\}}$
and $\Snp{a,b}{\exp\{ i \theta_k\}}$, where 
\begin{equation}
\theta_k = \tfrac{\pi}{3} k \;, \quad k=1,2 \;,
\label{theta}
\end{equation}
and  $\Li{a}{z}$ and $\Snp{a,b}{z}$ are the classical and 
Nielsen polylogarithms, respectively;
(iii)
the  additional statement that these constants should form an algebra.
This construction is in full agreement with 
the all-order $\ep$ expansions of the Feynman diagrams considered in \cite{ls}. 
In \cite{DK01}, this set of constants was extended up to weight ${\bf 5}$,
and it was shown that new transcendentals are generated 
($\chi_5$ in the notation of \cite{DK01})
which are not expressible in terms of Nielsen polylogarithms. 
Furthermore, the appearance  of the factor $\frac{1}{\sqrt{3}}$
only receives a strong explanation in \cite{DK04,MKL,series}.
This  factor comes from (i) the evaluation of a special type of multiple sums
and/or (ii) the structures of the coefficients of the all-order $\ep$ expansions 
of basic hypergeometric functions. In fact, we have\footnote{The structures of other types of sums 
and/or hypergeometric functions
were discussed in \cite{series,bin-sum,nested}.} reads:  
\begin{eqnarray} 
&& \hspace{-7mm}
F\left(\begin{array}{c|}
\{ 1+a_i\ep\}^{P}  \\
\tfrac{3}{2} \!+\! f\ep, \; 
\{ 1\!+\!e_i\ep \}^{P-2}
\end{array} ~z \right)
= \frac{(1 \!+\! 2 f\ep)}{2z}
\nonumber \\ && \hspace*{-7mm}
\times 
\frac{1\!-\!y}{1\!+\!y}
\left[ \ln y \!+\! \sum_{k=1}^\infty \ep^k \tilde{\tilde{\tilde{\Phi}}}_{k+1}(y)  \right]
\;,
\nonumber \\ && \hspace*{-7mm}
F\left(\begin{array}{c|}
\{ 1\!+\!a_i\ep\}^{R+2}, \; 
\{ 2\!+\!d_i\ep\}^{P\!-\!2\!-\!R}  \\
\tfrac{3}{2} \!+\! f\ep, \; 
\{ 1\!+\!e_i\ep \}^{R},
\{ 2\!+\!c_i\ep \}^{P\!-\!2\!-\!R} 
\end{array} ~z \right)
\nonumber \\ && \hspace*{-7mm}
= \frac{(1 \!+\! 2 f\ep)}{2z}
\Pi_{s=1}^{P\!-\!2\!-\!R\!} 
\frac{(1\!+\!c_s\ep)}  
     {(1\!+\!d_s\ep)}
\\ && \hspace*{-7mm}
\left\{ 
\frac{1\!-\!y}{1\!+\!y}
\left[ 
\ln  y \!+\! \sum_{k=1}^\infty \ep^k \Phi_{k\!+\!1}(y)
\right] 
+ \sum_{k=1}^\infty \ep^k \tilde{\Phi}_{k\!+\!1}(y)
\right\} \;,\ 
\label{HYPER1}
\nonumber \\
&& \hspace{-7mm}
F\left(\begin{array}{c|}
\{ 1\!+\!a_i\ep\}^{K+3}, \; \{ 2\!+\!d_i\ep\}^{P\!-\!3\!-\!K}  \\
\tfrac{3}{2} \!+\! f\ep, \; 
\{ 1\!+\!e_i\ep \}^{K\!-\!L}, \{ 2\!+\!c_i\ep \}^{P\!-\!2\!-\!K\!+\!L} 
\end{array} ~z \right)
\nonumber \\ && \hspace*{-7mm}
= \frac{(1 \!+\! 2 f\ep)}{2z}
\frac{\Pi_{s=1}^{P\!-\!2\!-\!K\!+\!L} (1\!+\!c_s\ep)}  
     {\Pi_{s=1}^{P\!-\!3\!-\!K} (1\!+\!d_s\ep)}
\sum_{k=0}^\infty \ep^k \tilde{\tilde{\Phi}}_{L+2+k}(y)\;,\ 
\label{HYPER2}
\end{eqnarray}
where 
$F$ is the hypergeometric function ${}_{P}F_{P-1}$, 
$R$, $K$ and $L$ are integers with 
$0 \leq R \leq P\!-\!2$, 
$0 \leq K \leq P\!-\!3$, 
$0 \leq L \leq K$,
the superscripts $R$ and $K-L$ indicate the lengths of the parameter lists,
$
y = \frac{1-\sqrt{\frac{z}{z-1}}}{1+\sqrt{\frac{z}{z-1}}},
$
and $\Phi_k(y), \tilde{\Phi}_k(y), \tilde{\tilde{\Phi}}_k(y)$ 
and 
$\tilde{\tilde{\tilde{\Phi}}}_{k+1}(y)$
are linear combinations of 
multiple polylogarithms of the square root of unity \cite{Goncharov}
of weight ${\bf k}$,
\begin{eqnarray}
\Phi_k(y)
&=& 
\sum_{\vec{s},j} 
c_{\vec{s}, k}
\ln^{j}(y) 
\left[
\Li{\left( \vec{\sigma} \atop \vec{s} \right)}{\pm y}   
\!-\! 
\Li{\left( \vec{\sigma} \atop \vec{s} \right)}{\pm 1} 
\right]\;.
\nonumber\\
&&\label{Phi}
\end{eqnarray}
Here, $c_{\vec{s}, k}$ are numerical coefficients, 
$\vec{s}=(s_1, \cdots s_n)$ and 
$\vec{\sigma} = (\sigma_1, \cdots, \sigma_n)$
are multi-indices, 
$\sigma_k = \pm 1$ are the square roots of unity,
$j+s_1+\cdots+s_n=k$, 
and  we have used the definition: 
$$
\Li{\left( \sigma_1, \cdots, \sigma_k \atop s_1, \cdots, s_n \right)}{z} 
= 
\sum_{m_1 > \cdots m_n > 0} \frac{z^{m_1} \sigma_1^{m_1} \cdots \sigma_n^{m_n}
                                            }{m_1^{s_1} m_2^{s_2} \cdots m_n^{s_n}} \;.
$$
The r.h.s.\ of Eq.~(\ref{Phi}) corresponds to the Remiddi-Vermaseren functions \cite{RV}. 
The differential-reduction algorithm \cite{MKL,Takayama,BKK} 
allows us to write the expansions for arbitrary values of parameters, as
\begin{eqnarray}
&& \hspace{-5mm}
Q(z)
_{P}F_{P-1}\left(\begin{array}{c|}
\{ I_i\!+\!a_i\ep\}^{P}  \\
L\!+\!\tfrac{1}{2} \!+\! f\ep, \; 
\{ K_i\!+\!e_i\ep \}^{P\!-\!2} 
\end{array} ~z \right)
\nonumber \\ && \hspace{-5mm}
= 
\sum R_i(z) [{\mbox {Eq}.~(\ref{HYPER1})}]
\!+\! 
\sum S_i(z) [{\mbox {Eq}.~(\ref{HYPER2})}]  \;,
\label{reduction}
\end{eqnarray}
where 
$Q(z)$, $R_i(z)$ and $S_i(z)$ 
are polynomials and 
$I_i$, $K_i$ and $L$ are integers. 
For two-cut diagrams with massive lines, 
the on-mass-shell case corresponds to $z=1/4$, so that 
$y = \exp(\pm i   \pi/3)$, 
$-y = \exp(\mp i 2 \pi/3)$ and 
$\frac{1-y}{1+y} = \mp \frac{i}{\sqrt{3}}$,
and the $\ep$ expansions of Eqs.~(\ref{HYPER1}) and (\ref{HYPER2}) have the structures
$
\frac{i}{\sqrt{3}}
\sum_{k=0}^\infty \ep^j \Phi_{j+1}\left( \pm \theta_k \right)
$
and 
$
\sum_{k=0}^\infty \ep^j \tilde{\Phi}_{L+1+j} \left( \pm \theta_k \right),
$
respectively, where 
$
\Phi_{j}(z)$ and $\tilde{\Phi}_j(z)$
are defined by Eq.~(\ref{Phi}), 
and $\theta_k$ by Eq.~(\ref{theta}).

{\bf 2}.
From the all-order $\ep$ expansions of the hypergeometric functions 
constructed in \cite{series}, the following set of 
constants are generated at weight {\bf k}: \\
(i)
The Remiddi-Vermaseren functions evaluated with arguments 
$\theta_k$  defined by Eq.~(\ref{theta}),
\begin{eqnarray}
\sum_{\vec{s},j} 
\left( i \pi \right)^j 
\left[
\Li{\left( \vec{\sigma} \atop \vec{s} \right)}{ \pm \theta_k }
\!-\! 
\Li{\left( \vec{\sigma} \atop \vec{s} \right)}{ \pm 1} \right] \;. 
\label{set1}
\end{eqnarray}
This is a subset of the ``sixth root of unity'' \cite{B99}, which, up to weight ${\bf 4}$, 
agrees with \cite{FK99} and, at weight {\bf 5}, with \cite{DK01}. 
(ii)
The product of Remiddi-Vermaseren functions, with arguments
$\theta_k$  defined by Eq.~(\ref{theta}), multiplied by $i/\sqrt{3}$,
\begin{eqnarray}
\frac{i}{\sqrt{3}}
\sum_{\vec{s},j} 
\left( i \pi \right)^j 
\left[
\Li{\left( \vec{\sigma} \atop \vec{s} \right)}{ \pm \theta_k }
\!-\! 
\Li{\left( \vec{\sigma} \atop \vec{s} \right)}{\pm 1} \right] \;. 
\label{set2}
\end{eqnarray}
Up to weights {\bf 4} and {\bf 5}, these results agree with \cite{FK99}  and \cite{DK01},
respectively. This explains the mysterious generation of the factor $1/\sqrt{3}$.

The product of hypergeometric functions can be understood as the product 
of low-loop master integrals (the 2-loop propagator diagram may generate 
the product of two 1-loop self-energies). 
The appearance of diagrams of this type gives rise exactly to the product 
of two basis elements described in \cite{FK99}. 

For practical applications in the PSLQ analysis \cite{pslq}, 
it is desirable to have a minimal set of 
constants corresponding to Eqs.~(\ref{set1}) and (\ref{set2}).
Their special subclass, Euler-Zagier sums, was
analyzed in \cite{EZ}.  
For the other constants, the solution is not unique, since 
the commonly accepted parametrizations
of the real and imaginary parts of the Remiddi-Vermaseren 
functions\footnote{%
For numerical evaluations of Remiddi-Vermaseren functions, some of the
existing programs \cite{program} may be used.}  
with argument $z=\exp(i \phi)$ do not exist.
For example, starting from weight {\bf 3}, the  {\it inverse binomial sums},
$
\sum_{n=1}^\infty  
\frac{1}{\left(2j \atop j\right)} \frac{1}{j^c},
$
can be written (for details, see \cite{DK01,DK00,more-sums,KV99})
in terms of either Dirichlet's $L$ series,   
the derivatives of $\Psi$ functions, 
generalized log-sin functions \cite{KV99}, 
Nielsen polylogarithms  \cite{DK00,DK01}
or generalized polylogarithms.
Let us recall that the classical polylogarithms with these arguments produce the
Clausen functions $\Cl{j}{\theta}$~\cite{Lewin},   
while the Nielsen polylogarithms produce 
the generalized log-sine functions $\LS{a}{k}{\theta}$ only
\cite{DK01,DK00,Lewin,review,lsjk}. 
For the parametrizations of Remiddi-Vermaseren functions with complex unit,
$\Lsc{a,b}{\theta}$ and $\LsLsc{a}{b}{c}{\theta}$ functions 
were introduced in \cite{DK04,review,three}. 
\\
{\bf 3.}
The $\ep$ expansions of hypergeometric functions 
entering the r.h.s.\ of Eqs.(\ref{HYPER1}) and (\ref{HYPER2})
may be written in terms of {\it multiple  inverse binomial sums}
\cite{DK01,DK04,DK00,KV99} defined as
\begin{equation}
\Sigma_{a_1,\ldots,a_p; \; b_1,\ldots,b_q;c}
\equiv
\sum_{j=1}^\infty 
\frac{
S_{a_1} \ldots S_{a_p} \Lambda_{b_1} \ldots \Lambda_{b_q} 
}{
\left( 2j \atop j\right) 
j^c}\;,
\label{binsum}
\end{equation}
where $S_a$ and $\bar{S}_b$ stand for $S_a(j-1)$ and $S_b(2j-1)$, respectively,
$S_k(j) = \sum_{l=1}^j l^{-k}$ are the harmonic sums and 
$\Lambda_a$ denotes the following linear combinations of
$\bar{S}_a$: 
\begin{eqnarray}
&& \hspace{-5mm}
\exp \left( \sum_{k=1}^\infty \ep^k \bar{S}_k \right) 
= 1 
+ \ep \bar{S_1}
+ \ep^2 \left[
\bar{S_2} \!+\! \bar{S}_1^2 
\right]
\nonumber \\ && \hspace{-5mm}
+ \ep^3 \left[
\bar{S_1}^3 \!+\! 3 \bar{S}_1 \bar{S}_2 \!+\! 2 \bar{S}_3 
\right]
\nonumber \\ && \hspace{-5mm}
+ \ep^4 \left[
\bar{S_1}^4 
\!+\! 6 \bar{S}_1^2 \bar{S}_2 
\!+\! 3 \bar{S}_2^2 
\!+\! 8 \bar{S}_1 \bar{S}_3 
\!+\! 6 \bar{S}_4 
\right]
\!+\! {\cal O} (\ep^5) 
\nonumber \\ && \hspace{-5mm}
= 
1+ \sum_{j=1}^\infty \ep^j \Lambda_j \;.
\nonumber 
\end{eqnarray}
Here, all rational factors coming from the series expansion of the exponent 
are equal to $1$.
For the hypergeometric functions of Eq.(\ref{HYPER1}),
we have
\begin{equation}
\sum_{j=1}^\infty 
\frac{
\Omega_r^{(K)}
}{
\left( 2j \atop j\right) j}
= \frac{1}{\sqrt{3}}
\sum_{r=1}^{m_\omega} c_r \omega_{r}^{(K+1)}\;, 
\label{basis1}
\end{equation}
where 
$c_r$ are rational numbers,
$\Omega_r^{(K)}$ are products of the harmonic sums $S_a$ and $\Lambda_b$, namely
\begin{eqnarray}
&& \hspace{-5mm}
\Omega_r^{(0)} = 1\;,\quad 
\Omega_r^{(1)} \in \{ S_1, \Lambda_1 \}\;, \quad  
\nonumber \\ && \hspace{-5mm}
\Omega_r^{(2)} \in \{ S_2, S_1^2, S_1 \Lambda_1, \Lambda_2 \}\;, 
\nonumber \\ && \hspace{-5mm}
\Omega_r^{(3)} \in  \{ S_3, S_1 S_2, S_1^3, 
S_2 \Lambda_1, 
S_1^2 \Lambda_1, 
S_1 \Lambda_2, 
\Lambda_3  \} \;, 
\nonumber \\ && \hspace{-5mm}
\Omega_r^{(4)} \in 
\{ S_4, S_1 S_3, S_2^2, S_1^2 S_2, S_1^4, 
\Lambda_4, 
S_1 \Lambda_3,  
\nonumber \\ && \hspace{5mm}
S_2 \Lambda_2, 
S_1^2 \Lambda_2, 
S_1^3 \Lambda_1,  
S_1 S_2 \Lambda_1, 
S_3 \Lambda_1 
\}\;,  
\end{eqnarray}
and $\omega_{r}^{(K)}$ belong to sets of transcendental
constants, namely 
\begin{eqnarray}
&& \hspace{-5mm}
\omega_r^{(1)} = \pi \;,\quad 
\omega_{r}^{(2)} \in \{ \Ls{2}{\tfrac{\pi}{3}},  \pi \ln 3, \}\;, 
\nonumber \\ && \hspace{-5mm}
\omega_{r}^{(3)} \in \{C_3, \pi \ln^2 3 , \pi \zeta_2 \} \;, 
\nonumber \\ && \hspace{-5mm}
\omega_{r}^{(4)} \in \{ 
C_4,
\Ls{4}{\tfrac{\pi}{3}},
\pi \zeta_3, 
\pi  \ln^3 3, 
\zeta_2 \omega_{r}^{(2)}
\}\;. 
\label{omega1234}
\end{eqnarray}
At weight {\bf 5}, there are 10 independent terms: 
\begin{eqnarray}
\omega_{1}^{(5)}  & = & C_5 \;, 
\omega_{2}^{(5)}  = D_1 \;, \quad 
\omega_{3}^{(5)}  = \Ls{5}{\tfrac{\pi}{3}}  \;, 
\nonumber \\ 
\omega_{4}^{(5)}  & = & \pi \zeta_4 \;,\ 
\omega_{5}^{(5)}  = \pi \zeta_3 \ln 3 \;,\ 
\omega_{6}^{(5)}  = \pi \zeta_2 \ln^2 3 \;, 
\nonumber \\ 
\omega_{7}^{(5)}  & = & \pi \ln^4 3 \;,  \quad 
\omega_{8}^{(5)}    = \pi \left[\Ls{2}{\tfrac{\pi}{3}} \right]^2\;, 
\nonumber \\ 
\omega_{9}^{(5)}  & = & \zeta_3 \Ls{2}{\tfrac{\pi}{3}} \;,\quad 
\omega_{10}^{(5)}  = \zeta_2 C_3 \;,
\label{omega}
\end{eqnarray}
where 
$\LS{j}{k}{\theta}$ are generalized log-sine functions defined as 
\begin{eqnarray}
\LS{j}{k}{\theta} & = &    - \int_0^\theta {\rm d}\phi \;
   \phi^k \ln^{j-k-1} \left| 2\sin\tfrac{\phi}{2}\right| \;,
\nonumber \\  
\Ls{j}{\theta} & = & \LS{j}{0}{\theta} \, ,
\label{log-sine}
\end{eqnarray}
and $C_3$, $C_4$, $C_5$ and $D_1$ are combinations of generalized log-sine functions, namely 
\begin{eqnarray}
C_3 & = &  3 \Ls{3}{\tfrac{2\pi}{3}} \!-\!  2 \Ls{2}{\tfrac{\pi}{3}} \ln 3 \;, 
\nonumber \\ 
C_4 & = & 
      2 \Ls{4}{\tfrac{2\pi}{3}}
\!-\! 3 \Ls{3}{\tfrac{2 \pi}{3}} \ln 3 
\!+\!   \Ls{2}{\tfrac{\pi}{3}} \ln^2 3 \;,
\nonumber \\ 
C_5 & = & 
\Ls{5}{\tfrac{2\pi}{3}}
\!-\! 2 \Ls{4}{\tfrac{2\pi}{3} } \ln 3
\nonumber \\ && 
+ \frac{3}{2} \Ls{3}{\tfrac{2\pi}{3}} \ln^2 3 
\!-\! \tfrac{1}{3} \Ls{2}{\tfrac{\pi}{3}} \ln^3 3 \;,
\nonumber \\ 
D_1 & = & 
3  \LS{5}{2}{\tfrac{2 \pi}{3}} \!-\! 4 \pi \LS{4}{1}{\tfrac{2 \pi}{3}} 
\nonumber \\ && 
+ \frac{32}{27} \Ls{4}{\tfrac{\pi}{3}} \ln 3 
+ 8 \zeta_2 \Ls{3}{\tfrac{2\pi}{3}}
\;.  
\label{CD}
\end{eqnarray}
We wish to mention that the combinations $C_3$, $C_4$ and $C_5$ completely coincide with 
appropriate terms of the $\ep$ expansions of the Gauss hypergeometric functions considered in \cite{ls},
and the generating function for these combinations is 
$$
\sum_{j=0}^\infty \ep^j C_{j+1} 
= \frac{3}{2} 3^{-\ep} \sum_{j=0}^\infty \frac{(2 \ep)^j}{j!} \Ls{j+1}{\frac{2}{3} \pi} \;,
$$
so that $C_1=0$ and $C_2 = 2 \Ls{2}{\tfrac{\pi}{3}}$.
For the hypergeometric functions of Eq.(\ref{HYPER2}), the $\ep$ expansions have the form
\begin{equation}
\sum_{j=1}^\infty 
\frac{
\Omega_k^{(R)}
}{
\left( 2j \atop j\right) j^{2+r}}
= 
\sum_{s=1}^{L_\sigma} C_s \sigma_{s}^{(2+r+R)}\;, 
\label{basis2}
\end{equation}
where $r$ is a positive integer, 
$C_s$ is a rational number
and $\sigma_{r}^{(K)}$ belong to sets of transcendental constants, namely 
\begin{eqnarray}
&& 
\sigma_r^{(2)} = \zeta_2 \;,\quad 
\sigma_{r}^{(3)} \in \{ \zeta_3, \pi \Ls{2}{\tfrac{\pi}{3}} \}\;, 
\nonumber \\ && 
\sigma_{r}^{(4)} \in \{ \pi \Ls{3}{\tfrac{2\pi}{3}}, \left[  \Ls{2}{\tfrac{\pi}{3}} \right]^2, \zeta_4 \} \;. 
\label{sigma1234}
\end{eqnarray}
At weight {\bf 5}, there are only 6 independent constants, namely 
\begin{eqnarray}
\sigma_{1}^{(5)} & = &  \pi \Ls{4}{\tfrac{\pi}{3}} \;,\quad
\sigma_{2}^{(5)} =  \pi \Ls{4}{\tfrac{2\pi}{3}} \;,  
\nonumber \\  
\sigma_{3}^{(5)} & = &  \pi \zeta_2 \Ls{2}{\tfrac{\pi}{3}} \;,\quad   
\sigma_{4}^{(5)} = \chi_5 \;, 
\nonumber \\  
\sigma_{5}^{(5)} & = & \zeta_5 \;, \quad
\sigma_{6}^{(5)} = \zeta_2 \zeta_3 \;, 
\label{sigma}
\end{eqnarray}
where $\chi_5$ is defined as
$\chi_5 \equiv \sum_{j=1}^\infty \frac{S_1^3}{\left( 2j \atop j\right) j^2}$ \cite{DK01,review}.
For illustration, we present here the analytical results for 
all sums of weight {\bf 5}:
\begin{eqnarray}
&&
\left \langle \frac{S_4}{j}  \right \rangle  = 
      \frac{4}{9} \omega_3^{(5)}
\!+\! \frac{155}{54} \pi \zeta_4
\!-\! \frac{2}{3} \omega_8^{(5)}
\!+\! \frac{8}{3} \omega_9^{(5)}
\;,
\nonumber \\ && 
\left \langle  \frac{S_2^2}{j}  \right \rangle   
 = 
      \frac{4}{9} \omega_3^{(5)}
\!+\! \frac{233}{81} \pi \zeta_4
- \frac{2}{3} \omega_8^{(5)}
+ \frac{8}{3} \omega_9^{(5)}
\;,  
\nonumber \\ && 
\left \langle  \frac{S_1 S_3}{j}  \right \rangle   
 = 
- \frac{9}{8} D_1
\!+\! \frac{107}{54} \omega_3^{(5)}
\!+\!  \frac{2437}{432} \pi \zeta_4
\nonumber \\ && \hspace{5mm}
{}+ \frac{16}{9} \omega_5^{(5)}
\!+\! \frac{2}{9} \omega_8^{(5)}
\!-\! \frac{14}{9} \omega_9^{(5)}
\!-\! \frac{1}{3} \omega_{10}^{(5)} 
\;,
\nonumber \\ && 
\left \langle  \frac{S_1^2S_2}{j}  \right \rangle   
 = 
       \frac{5}{12} D_1
\!-\!  \frac{131}{81} \omega_3^{(5)}
\!-\! \frac{1843}{216} \pi \zeta_4
\nonumber \\ && \hspace{5mm}
{}- \frac{98}{81} \omega_5^{(5)}
+ \frac{1}{27} \omega_6^{(5)}
\!+\! \frac{14}{27} \omega_8^{(5)}
\!-\! \frac{68}{27} \omega_9^{(5)}
\nonumber \\ && \hspace{5mm}
{}\!-\! \frac{2}{9} \omega_{10}^{(5)} 
\;,
\nonumber \\ && 
\left \langle  \frac{S_1^4}{j}  \right  \rangle   
 = 
16 C_5 
\!+\! \frac{23}{2} D_1
\!-\!  \frac{218}{9} \omega_3^{(5)}
\!+\! \frac{2837}{108} \pi \zeta_4
\nonumber \\ && \hspace{5mm}
{}+ \frac{716}{27} \omega_5^{(5)}
\!+\! \frac{110}{9} \omega_6^{(5)}
\!+\! \frac{1}{3} \omega_7^{(5)}
\!-\! \frac{2}{3} \omega_8^{(5)}
\nonumber \\ && \hspace{5mm}
{}+ \frac{16}{3} \omega_9^{(5)}
\!+\! \frac{4}{3} \omega_{10}^{(5)} 
\;,
%
%
\nonumber \\ && 
\left \langle  \frac{S_1^3 \Lambda_1}{j} \right \rangle   
 = 
19 C_5 
\!+\! \frac{195}{16} D_1
\!-\!  \frac{2717}{108} \omega_3^{(5)}
\nonumber \\ && \hspace{5mm}
{}+ \frac{40093}{864} \pi \zeta_4
+ \frac{302}{9} \omega_5^{(5)}
\!+\! \frac{89}{6} \omega_6^{(5)}
\!+\! \frac{1}{3} \omega_7^{(5)}
\nonumber \\ && \hspace{5mm}
{}- \frac{13}{9} \omega_8^{(5)}
\!+\! \frac{94}{9} \omega_9^{(5)}
\!+\! \frac{5}{3} \omega_{10}^{(5)} 
\;,
\nonumber \\ && 
\left \langle  \frac{S_1 S_2 \Lambda_1}{j} \right \rangle   
 = 
\frac{83}{48} D_1
\!-\!  \frac{1493}{324} \omega_3^{(5)}
\nonumber \\ && \hspace{5mm}
{}- \frac{50435}{2592} \pi \zeta_4
\!-\! \frac{278}{81} \omega_5^{(5)}
\!+\! \frac{1}{27} \omega_6^{(5)}
\nonumber \\ && \hspace{5mm}
{}+ \frac{17}{27} \omega_8^{(5)}
\!-\! \frac{74}{27} \omega_9^{(5)}
\!+\! \frac{1}{9} \omega_{10}^{(5)} 
\;,
\nonumber \\ && 
\left[ \frac{S_3 \Lambda_1}{j} \right \rangle   
 = 
- \frac{9}{8} D_1
\!+\!  \frac{89}{54} \omega_3^{(5)}
\nonumber \\ && \hspace{5mm}
{}+ \frac{1855}{432} \pi \zeta_4
\!+\! \frac{16}{9} \omega_5^{(5)}
\!+\! \frac{11}{9} \omega_8^{(5)}
\nonumber \\ && \hspace{5mm}
{}- \frac{62}{9} \omega_9^{(5)}
\!-\! \frac{1}{3} \omega_{10}^{(5)} 
\;,
\nonumber \\ &&
\left \langle  \frac{S_1^2 \Lambda_2}{j} \right \rangle   
 = 
22 C_5
\!+\! \frac{257}{24} D_1
\!-\!  \frac{3593}{162} \omega_3^{(5)}
\nonumber \\ && \hspace{5mm}
{}+ \frac{102433}{1296} \pi \zeta_4
\!+\! \frac{3470}{81} \omega_5^{(5)}
\!+\! \frac{482}{27} \omega_6^{(5)}
\nonumber \\ && \hspace{5mm}
{}\!+\! \frac{1}{3} \omega_7^{(5)}
\!-\! \frac{56}{27} \omega_8^{(5)}
\!+\! \frac{404}{27} \omega_9^{(5)}
\!+\! \frac{14}{9} \omega_{10}^{(5)} 
\;,
\nonumber \\ && 
\left \langle  \frac{S_2 \Lambda_2}{j} \right \rangle   
 = 
 \frac{73}{24} D_1
\!-\!  \frac{1249}{162} \omega_3^{(5)}
\nonumber \\ && \hspace{5mm}
{}- \frac{42319}{1296} \pi \zeta_4
\!-\! \frac{458}{81} \omega_5^{(5)}
\!+\! \frac{1}{27} \omega_6^{(5)}
\nonumber \\ && \hspace{5mm}
{}\!-\! \frac{16}{27} \omega_8^{(5)}
\!+\! \frac{136}{27} \omega_9^{(5)}
\!+\! \frac{4}{9} \omega_{10}^{(5)} 
\;,
\nonumber \\ && 
\left\langle  \frac{S_1 \Lambda_3}{j} \right \rangle   
 = 
25 C_5 
\!+\! \frac{61}{8} D_1
\!-\! \frac{949}{54} \omega_3^{(5)}
\nonumber \\ && \hspace{5mm}
{}+ \frac{16771}{144} \pi \zeta_4
\!+\! \frac{1402}{27} \omega_5^{(5)}
\!+\! \frac{383}{18} \omega_6^{(5)}
\nonumber \\ && \hspace{5mm}
{}+ \frac{1}{3} \omega_7^{(5)}
\!-\! \frac{16}{9} \omega_8^{(5)}
\!+\! \frac{124}{9} \omega_9^{(5)}
\!+\! \frac{4}{3} \omega_{10}^{(5)} 
\;,
\nonumber \\ && 
\left\langle  \frac{\Lambda_4}{j} \right \rangle   
 = 
28 C_5 
\!+\! \frac{7}{2} D_1
\!-\! \frac{422}{27} \omega_3^{(5)}
\nonumber \\ && \hspace{5mm}
{}+ \frac{5093}{36} \pi \zeta_4
\!+\! \frac{1576}{27} \omega_5^{(5)}
\!+\! \frac{226}{9} \omega_6^{(5)}
\nonumber \\ && \hspace{5mm}
{}+ \frac{1}{3} \omega_7^{(5)}   
\!-\! \frac{16}{9} \omega_8^{(5)}
\!+\! \frac{136}{9} \omega_9^{(5)}
\!+\! \frac{4}{3} \omega_{10}^{(5)} 
\;,
%
%
%
\\ && 
\left[ \frac{S_1}{j^4}  \right] = 
- \frac{28}{81}\sigma_{1}^{(5)} 
\!+\! \frac{19}{27} \zeta_2 \zeta_3 \!+\! \frac{134}{27} \zeta_5 \;,
\nonumber \\ && 
\left[ \frac{\Lambda_1}{j^4} \right] = 
- \frac{82}{81}\sigma_{1}^{(5)}
\!+\! \frac{46}{27} \zeta_2 \zeta_3 
\!+\! \frac{847}{54} \zeta_5 \;,
\nonumber \\ && 
\left[ \frac{S_3}{n^2} \right] = 
- \frac{4}{9}\sigma_{1}^{(5)} 
\!+\! \frac{8}{9} \zeta_2 \zeta_3
\!+\! \frac{58}{9} \zeta_5 \;,
\nonumber \\ && 
\left[ \frac{S_2}{n^3} \right] = 
- \frac{4}{27}\sigma_{1}^{(5)} 
\!+\! \frac{2}{27} \sigma_{3}^{(5)}  
\!+\! \frac{2}{3}  \zeta_2 \zeta_3 
\!+\! \frac{29}{27} \zeta_5 \;,
\nonumber \\ && 
\left[  \frac{S_1 S_2}{n^2} \right] = 
   \frac{20}{243}\sigma_{1}^{(5)} 
\!-\! \frac{4}{81} \sigma_{3}^{(5)} 
\nonumber \\ && \hspace{15mm}
\!-\! \frac{23}{81} \zeta_2 \zeta_3 
\!-\! \frac{53}{81} \zeta_5
\;,
\nonumber \\ && 
\left[\frac{S_2 \Lambda_1}{n^2}  \right] = 
 \frac{146}{243}\sigma_{1}^{(5)} 
\!-\! \frac{7}{81} \sigma_{3}^{(5)}
\nonumber \\ && \hspace{15mm}
\!-\! \frac{95}{81}  \zeta_2 \zeta_3 
\!-\! \frac{662}{81} \zeta_5 
\;,
\nonumber \\ &&
\left[ \frac{S_1^2}{n^3}  \right] = 
 \frac{2}{9}\sigma_{1}^{(5)} 
\!-\! \frac{1}{2} \chi_5  
\!-\! \frac{13}{18}  \zeta_2 \zeta_3 
\!-\! \frac{47}{18} \zeta_5 \;,
\nonumber \\ &&
\left[ \frac{S_1 \Lambda_1}{n^3}  \right] = 
 \frac{277}{324}\sigma_{1}^{(5)} 
\!-\! \frac{1}{2} \sigma_{2}^{(5)}
\!+\! \frac{23}{108} \sigma_{3}^{(5)}
\nonumber \\ && \hspace{5mm}
{}+ \frac{323}{432}  \zeta_2 \zeta_3 
\!-\! \frac{1291}{144} \zeta_5 
\!-\! \frac{11}{16} \chi_5  
\;,
\nonumber \\ && 
\left[\frac{S_1^2 \Lambda_1}{n^2}  \right] = 
- \frac{116}{243}\sigma_{1}^{(5)} 
\!+\! \frac{2}{3} \sigma_{2}^{(5)}
\!-\! \frac{23}{81} \sigma_{3}^{(5)}
\nonumber \\ && \hspace{5mm}
{}- \frac{437}{162}  \zeta_2 \zeta_3 
\!+\! \frac{529}{162} \zeta_5 
\!+\! \frac{3}{2} \chi_5
\;,
\nonumber \\ && 
\left[ \frac{\Lambda_2}{n^3}  \right] =
 \frac{337}{162}\sigma_{1}^{(5)} 
\!-\! \sigma_{2}^{(5)}
\!+\! \frac{67}{54} \sigma_{3}^{(5)}
\nonumber \\ && \hspace{5mm}
{}+ \frac{335}{216}  \zeta_2 \zeta_3 
\!-\! \frac{6037}{216} \zeta_5 
\!-\! \frac{7}{8} \chi_5  
\;,
\nonumber \\ && 
\left[  \frac{\Lambda_3}{n^2} \right] =
- \frac{1015}{324}\sigma_{1}^{(5)} 
\!+\! \frac{7}{2} \sigma_{2}^{(5)}
\!-\! \frac{469}{108} \sigma_{3}^{(5)}
\nonumber \\ && \hspace{5mm}
{}- \frac{5633}{432}  \zeta_2 \zeta_3 
\!+\! \frac{19123}{432} \zeta_5 
\!+\! \frac{49}{16} \chi_5 
\;,
\nonumber \\ && 
\left[ \frac{S_1 \Lambda_2}{n^2}  \right] = 
- \frac{517}{324}\sigma_{1}^{(5)} 
\!+\! \frac{11}{6} \sigma_{2}^{(5)}
\!-\! \frac{143}{108} \sigma_{3}^{(5)}
\nonumber \\ && \hspace{5mm}
{}- \frac{3059}{432}  \zeta_2 \zeta_3 
\!+\! \frac{7049}{432} \zeta_5 
\!+\! \frac{35}{16} \chi_5 
\;,
\end{eqnarray}
where 
$
\left\langle  X  \right \rangle   
=  \sqrt{3} \sum_{n=1}^\infty \frac{1}{\left( 2j \atop j\right)} X \;, 
$
$
\left[ X \right] = \sum_{n=1}^\infty \frac{1}{\left( 2j \atop j\right)} X
$
and $C_5$, $D_1$, $\omega_r^{(5)}$ and $\sigma_r^{(5)}$ are defined by 
Eqs.~(\ref{omega}), (\ref{CD}) and (\ref{sigma}). 
%
%
%
%
%
%
\\
{\bf 4.}
Based on the recently established analytical structures of the coefficients of the
all-order $\ep$ expansions of hypergeometric functions, indicated by 
Eqs.~(\ref{HYPER1}), (\ref{HYPER2}) and (\ref{reduction}), 
we presented in Eqs.~(\ref{omega1234}), (\ref{omega}), (\ref{sigma1234}) and (\ref{sigma}) 
the set of linearly independent transcendental constants generated  by single-scale massive Feynman diagrams with two massive cuts 
in $4-2\ep$ dimensions. 
The main difference between these two sets is the factor $\frac{1}{\sqrt{3}}$, which was 
predicted  in \cite{FK99}. 
We mention that the first set, given in Eqs.~(\ref{omega1234}) and (\ref{omega}), 
does not form an algebra: $\omega_r^{(a)} \times \omega_r^{(b)} \neq \omega_r^{(a+b)}$.
On the other hand, the second set, given in Eqs.~(\ref{sigma1234}) and (\ref{sigma}), does form an algebra. 
Changing the space-time dimension or
considering Feynman diagrams related to hypergeometric functions of a few variables 
or with a more complicated combination of  massive and massless cuts  
generates another set of constants, as may be seen, for example, in \cite{review,three,vertex}. 

\noindent
{\bf Acknowledgments.}
MYK would like to thank the organizers of the conference 
{\it Loops and Legs in Quantum Field Theory} for the invitation and for creating such a stimulating
atmosphere.
We are grateful to Z.~Merebashvili for carefully reading this manuscript.
This research was supported in part by BMBF Grant No.\ 05~HT6GUA, 
by DFG Grants No.\ KN~365/3--1 and KN~365/3--2, and by HGF Grant No.\ HA~101.


\end{document}